\documentclass[twocolumn,twoside,slac_two]{revtex4}
\usepackage{graphicx}
\usepackage{fancyhdr}
\usepackage{hyperref}
\hypersetup{
    colorlinks=true,
    urlcolor=magenta,
}
\begin{document}

\title{Improvement of the GAMMA-400 physical scheme for precision gamma-ray emission investigations}

%

\author{A. A. Leonov, A. M. Galper, I.V. Arkhangelskaja, A.I. Arkhangelskiy, V.V. Kadilin, V.A. Kaplin, M.D. Kheymits, N.A. Glushkov, V.V. Mikhailov, P.Yu. Naumov, A.A. Taraskin, M.F. Runtso, Yu.T. Yurkin}
\affiliation{National Research Nuclear University MEPhI (Moscow Engineering Physics Institute), 115409 Moscow, Russia}

\author{N.P. Topchiev, O.D. Dalkarov, A.E. Egorov, Yu.V. Gusakov, Yu.I. Stozhkov, S.I. Suchkov, V.G. Zverev}
\affiliation{Lebedev Physical Institute, Russian Academy of Sciences, 119991 Moscow, Russia}

\author{V. Bonvicini, M. Boezio, F. Longo, E. Mocchiutti}
\affiliation{Istituto Nazionale di Fisica Nucleare, Sezione di Trieste,  I-34127 Trieste, Italy}

\author{A.V. Bakaldin, S.G. Bobkov, M.S. Gorbunov, O.V. Serdin}
\affiliation{Scientific Research Institute for System Analysis, Russian Academy of Sciences, 117218 Moscow,  Russia}

\author{O. Adriani, P. Spillantini}
\affiliation{Istituto Nazionale di Fisica Nucleare, Sezione di Firenze, 50019 Firenze, Italy}

\author{P. Picozza, R. Sparvoli}
\affiliation{Istituto Nazionale di Fisica Nucleare, Sezione di Roma 2 and Physics Department of University of Rome Tor Vergata, Rome, 00133 Rome, Italy}

\author{A.A. Moiseev}
\affiliation{CRESST/GSFC and University of Maryland, College Park, 20742 Maryland, USA}

\author{I.V. Moskalenko}
\affiliation{Hansen Experimental Physics Laboratory and Kavli Institute for Particle Astrophysics and Cosmology, Stanford University, 94305-4085 Stanford, USA}

\author{M. Tavani}
\affiliation{Istituto Nazionale di Astrofisica and Physics Department of University of Rome Tor Vergata, 00136 Rome, Italy}

\author{B.I. Hnatyk}
\affiliation{Taras Shevchenko National University,  01601 Kyiv, Ukraine}

\author{V.E. Korepanov}
\affiliation{Lviv Center of Institute of Space Research, 79060 Lviv, Ukraine}

\begin{abstract}
The main goal for the GAMMA-400 gamma-ray telescope mission is to perform a sensitive search for signatures of dark matter particles in high-energy gamma-ray emission. Measurements will also concern the following scientific goals: detailed study of the Galactic center region, investigation of point and extended gamma-ray sources, studies of the energy spectra of Galactic and extragalactic diffuse emissions. To perform these measurements the GAMMA-400 gamma-ray telescope possesses unique physical characteristics for energy range from $\sim$20 MeV to $\sim$1000 GeV in comparison with previous and current space and ground-based experiments. The major advantage of the GAMMA-400 instrument is excellent angular and energy resolutions for gamma-rays above 10 GeV. The gamma-ray telescope angular and energy resolutions for the main aperture at 100-GeV gamma rays are $\sim$0.01$^{\circ}$ and $\sim$1$\%$, respectively. The special goal is to improve physical characteristics in the low-energy range from $\sim$20 MeV to 100 MeV. Minimizing the amount of dead matter in the telescope aperture allows us to obtain the angular and energy resolutions better in this range than in current space missions. The gamma-ray telescope angular resolution at 50-MeV gamma rays is better than 5$^{\circ}$ and energy resolution is $\sim$10$\%$. We report the method providing these results.
\end{abstract}

\maketitle
\thispagestyle{fancy}

\section{INTRODUCTION}
The major advantage of the GAMMA-400 instrument is excellent angular and energy resolutions for gamma-rays above 10 GeV. The gamma-ray telescope angular and energy resolutions for the main aperture at 100-GeV gamma rays are $\sim$0.01$^{\circ}$ and $\sim$1$\%$, respectively. The motivation of presented results is to improve physical characteristics of the GAMMA-400 gamma-ray telescope in the energy range of $\sim$20-100 MeV. Minimizing the amount of dead matter in the telescope aperture allows us to obtain the angular resolution better than in the current space missions in this energy range. The gamma-ray telescope angular resolution for 50-MeV gamma rays is better than 5$^{\circ}$. The energy resolution with the presented construction of the gamma-ray telescope GAMMA-400 is about 10$\%$ for 50-MeV gamma rays.

\section{THE GAMMA-400 GAMMA-RAY TELESCOPE}

The GAMMA-400 physical scheme is shown in FIG. 1. From the top, the telescope consists of the following layers:
\begin{itemize}
\item	the anticoincidence system (AC) is composed by two-layer plastic scintillators, located both on top and on the lateral side of the apparatus. The system is essentially used to veto charged particles;
\item the converter-tracker system (C) consists of 22 layers. 20 layers of converter-tracker have high-Z material (tungsten), in which $\gammaγ$-rays incident on the instrument can convert to an e$^+$/e$^-$pair. The converter planes are interleaved with position-sensitive detectors that record the passage of charged particles, thus measuring the tracks of the particles resulting from pair conversion. The position-sensitive detectors are double (x, y) silicon strips (pitch 0.08 mm). The lowest two (x, y)-planes have no tungsten converter material. The total converter-tracker thickness is about $\sim$1X$_0$ (X$_0$ is the radiation length). The converter-tracker information is used to precisely determine the conversion point and the direction of each incident particle. Also this information provides the possibility to measure polarization of gamma-rays;
\item the time of flight system (ToF) is formed by plastic scintillators S1 and S2, separated by approximately 500 mm. This system is used both to generate the trigger for the apparatus and to reject albedo particles by measuring time of particle passage;
\item the deep electromagnetic calorimeter CC. The total calorimeter thickness is $\sim$21 X$_0$ or $\sim$1.0$\lambda$$_0$ (where $\lambda$$_0$ is nuclear interaction length). Using a deep calorimeter allows us to extend the energy range up to several TeV for gamma rays, and to reach an energy resolution of approximately 1$\%$ above 100 GeV;
\item the scintillation detector S3 improves hadrons and electromagnetic showers separation.
\end{itemize}

The main difference of this scheme from the previous one is using 20 thin layers of converter foils with thickness 0.025X$_0$, against 8 layers of tungsten with thickness 0.1X$_0$ [1].\\
 $\hspace*{0.4cm}$The GAMMA-400 gamma-ray observatory will be installed onboard the Navigator space platform, which is designed and manufactured by the Lavochkin Association [1].

\begin{figure}[h]
\includegraphics[width=22pc]{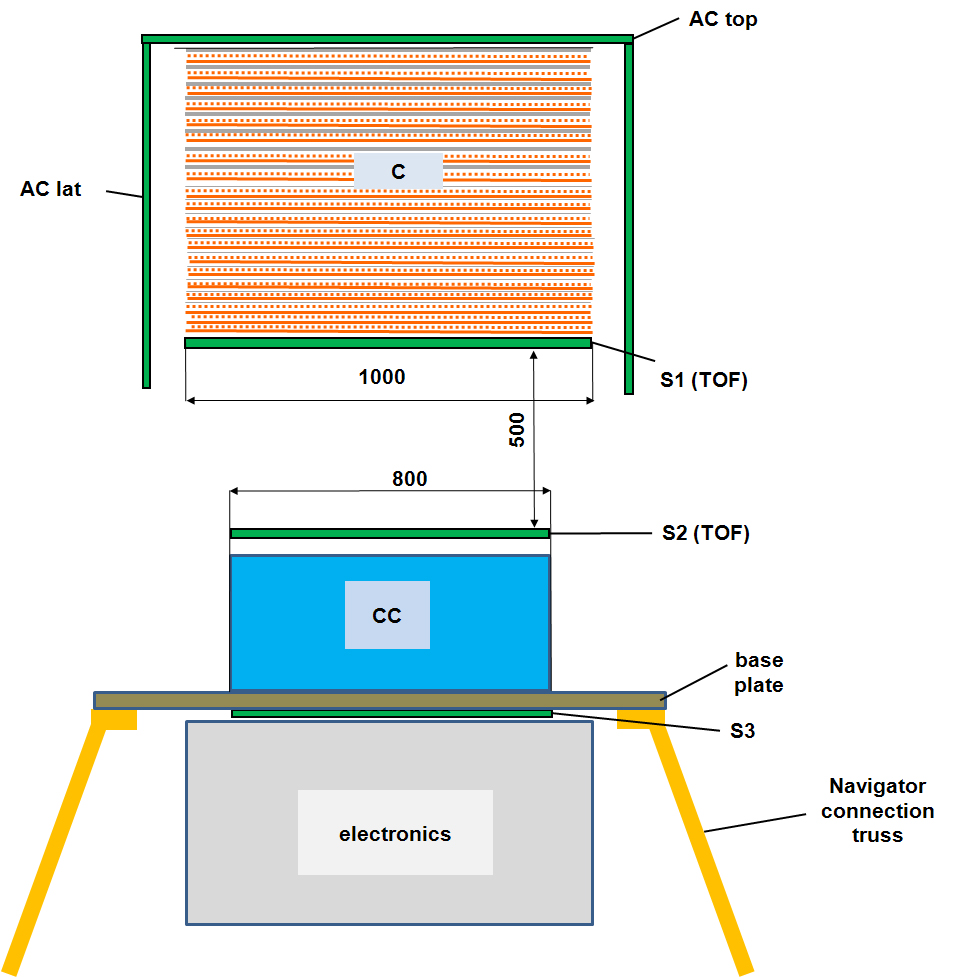}\hspace{2pc}%
\begin{minipage}[b]{22pc}\caption{\label{scheme}The GAMMA-400 physical scheme.}
\end{minipage}
\end{figure}

\section{METHOD TO RECONSTRUCT INCIDENT ANGLE FOR LOW ENERGY GAMMA RAYS}
The method to reconstruct the initial angle of incident low-energy gamma rays in the GAMMA-400 instrument was described in [2]. In this method, the effect of multiple scattering of produced pair components is used to involve the energy correction in the angle reconstruction procedure, involving the imaginary curvature radius for each component of the pair (FIG. 2).
When simulating the support structures for the detectors and for the converter foil planes were taken into account to check its influence on converter-tracker performance. We applied a similar construction as in the Fermi-LAT mission [3, 4]. The converter-tracker tower consists from the 23 trays supported by carbon-composite sidewalls with thickness of 0.8 mm. Each tray includes aluminum honeycomb core and has thickness about 3 cm. The total thickness of the material just from above each tungsten foil is about 0.01X$_0$ and comparable with thickness of tungsten layer 0.025X$_0$ for pair production (FIG. 2). The geometrical thickness of the matter above tungsten is 3 cm that two orders of magnitude more than geometrical thickness of converter foil (0.09 mm). For such structure, the accuracy of vertical localization of gamma-ray conversion point is quite different for the cases shown in FIG. 2.

\begin{figure}[h]
\includegraphics[width=20pc]{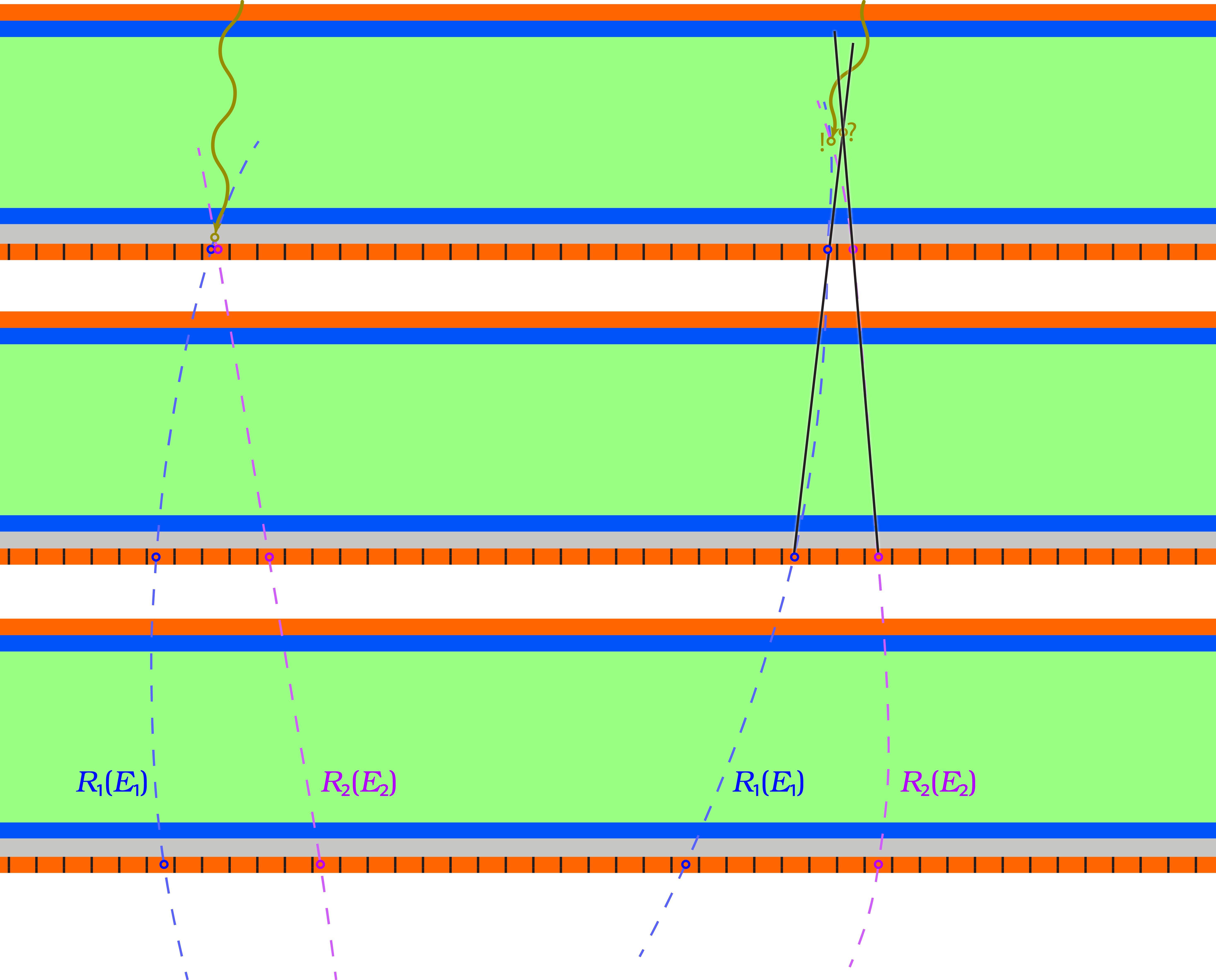}\hspace{0.01pc}%
\begin{minipage}[b]{20pc}\caption{\label{fork4}Two types of gamma-ray conversion events in converter-tracker. Orange: plane of silicon strips; blue: carbon-composite layers; green: aluminum honeycomb layers; grey: tungsten layers (0.025X$_0$). R$_i$(E$_i$) is the radius R$_i$ of imaginary curvature for the electron (positron) with the energy E$_i$ after conversion.}
\end{minipage}
\end{figure}

In the left part of FIG. 2, the conversion is occurred inside tungsten layer, and in the right part of FIG. 2, the conversion is appeared inside support matter. These two types of conversion events can be easily identified from the information of strip detectors in the tracker layer just under conversion point. In the “left” case, both pair components release energy in first single strip (1 point event), while in the “right” case, each component of the pair releases energy in separate strip (2 points event).
If one extract the events with 1-point topology, the accuracy of angular reconstruction appears significantly better than for events with 2-points topology. The results of angle reconstruction are shown in FIG. 3a  for incident gamma rays with energy of 50 MeV. The distributions for deviation angle between reconstructed direction and initial direction are shown. The initial gamma-ray direction was fixed as 2$^{\circ}$ for zenith angle and 45$^{\circ}$ for the polar angle just to check the robustness of the algorithm out from vertical direction. The angular resolution, defined as condition of 68$\%$ containment, for 1-point topology events is $\sim$4.6$^{\circ}$ and for 2-point topology events is $\sim$7$^{\circ}$.
The energy dependence of the angular resolution of the GAMMA-400 gamma-ray telescope is presented in FIG. 3b for 1-point topology, 2-point topology, and combined (1 point and 2 points) topology events. The angular resolution of Fermi-LAT instrument for on-axis gamma rays is also shown [3]. 

\begin{figure}[h]
\includegraphics[width=35pc]{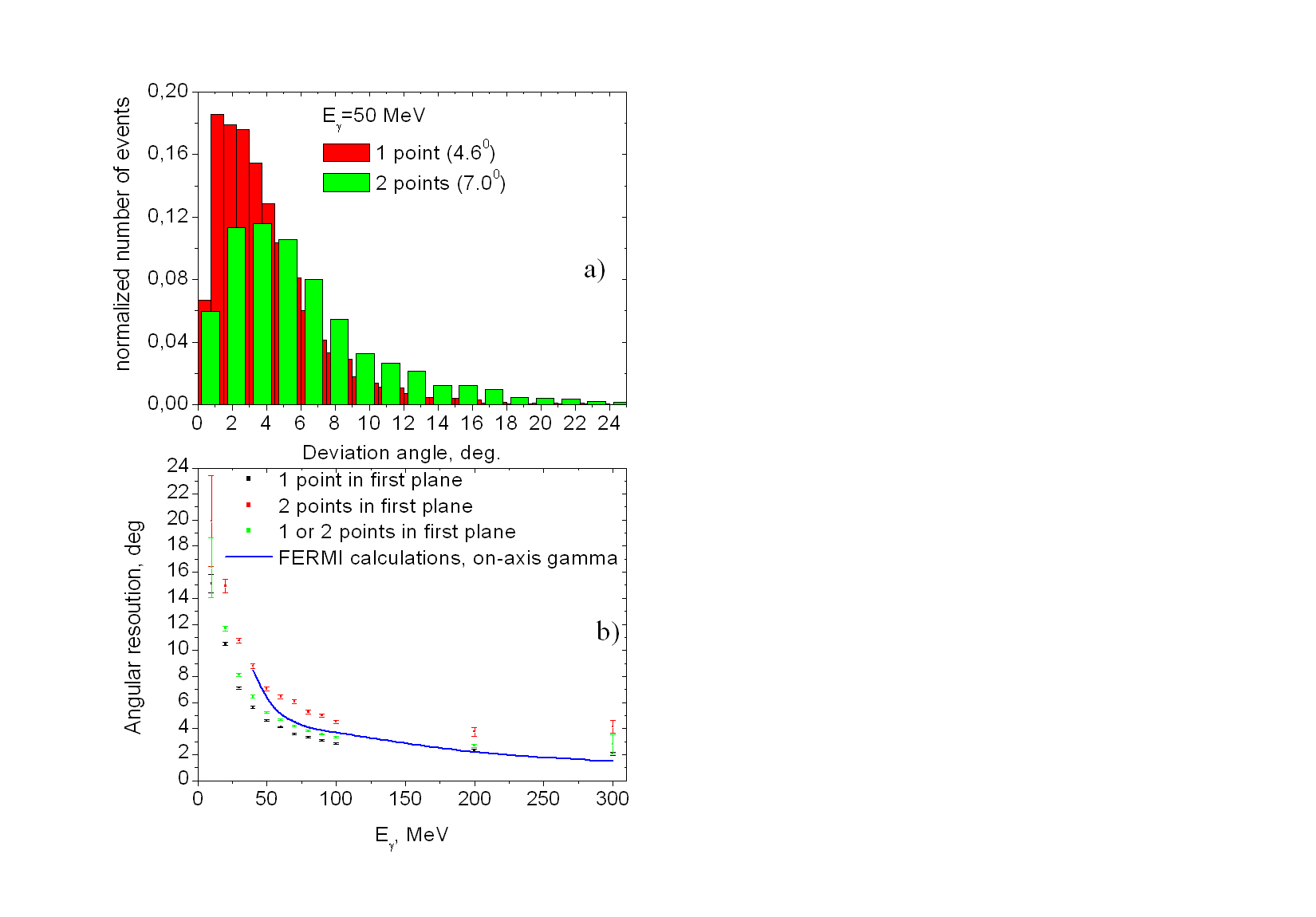}\hspace{2pc}%
\begin{minipage}[b]{20pc}\caption{\label{FIG3}(a) The distributions of the deviation angle between the reconstructed and initial direction for incident gamma rays with energy of 50 MeV. The angular resolution is determined as condition of 68$\%$ containment. (b) The energy dependence of the GAMMA-400 angular resolution (low energy branch). Black points: 1-point topology events; red points: 2 point-topology events; green points: 1 or 2 point topology events. The results of calculation of angular resolution for the Fermi-LAT on-axis gamma rays are shown by blue line (b).}
\end{minipage}
\end{figure}

The angular resolution for 1-point events in the GAMMA-400 converter-tracker is better than angular resolution obtained from Fermi-LAT simulation data for gamma-ray energy less, than 200 MeV. For gamma-ray energy range from 200 MeV to 1 TeV another method is used to reconstruct initial gamma-ray direction in the GAMMA-400 gamma-ray telescope [5]. The energy dependence of angular resolution of the GAMMA-400 gamma-ray telescope in the energy range from 10 MeV to 3 TeV is presented in FIG. 4a. PSF of the Fermi-LAT telescope for front configuration is also shown there [6]. The GAMMA-400 angular resolution  is significantly better than ones of Fermi-LAT, begining from the gamma-ray energy of 10 GeV.

\begin{figure}[h]
\includegraphics[width=35pc]{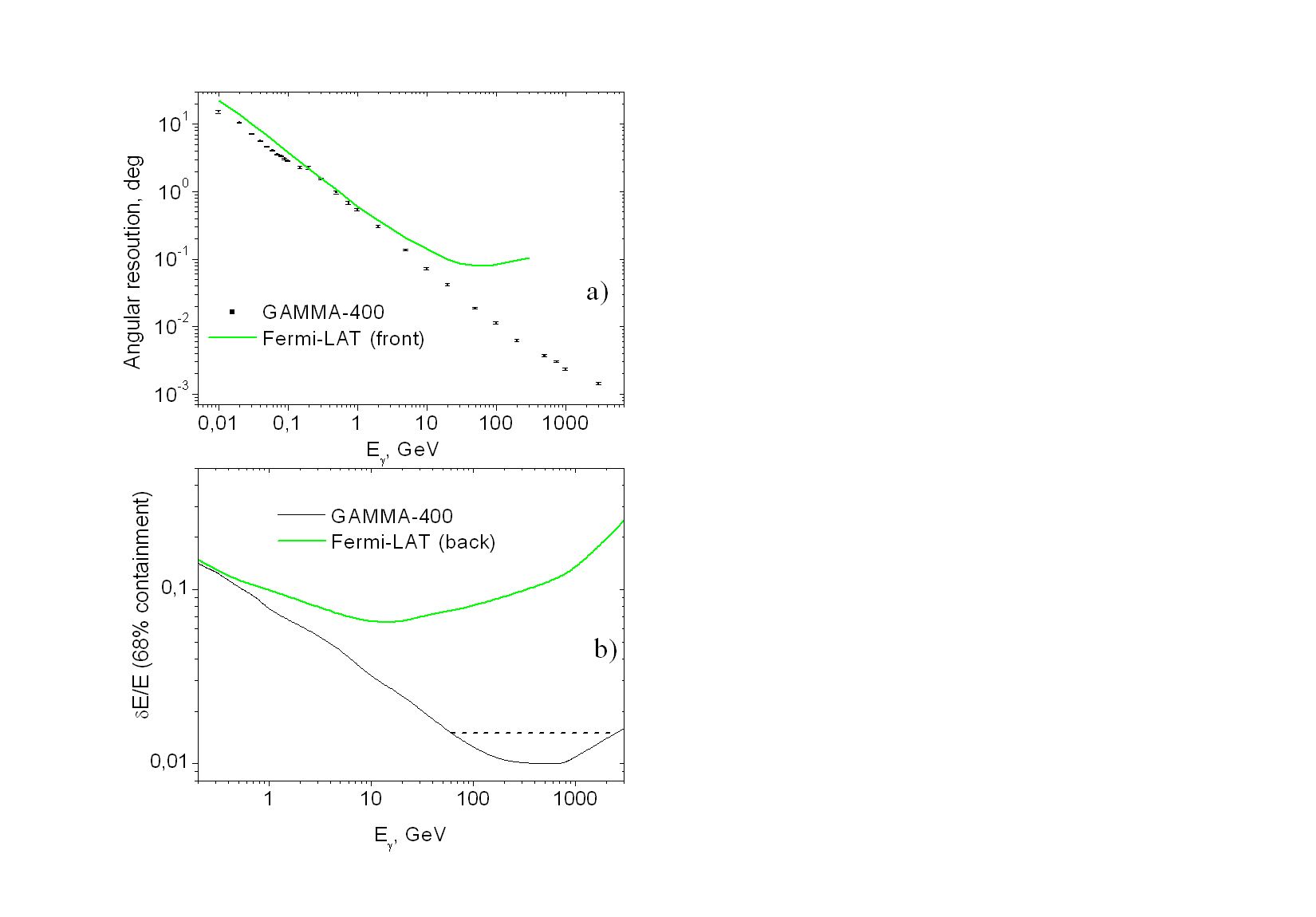}\hspace{2pc}%
\begin{minipage}[b]{20pc}\caption{\label{FIG4}(a) The energy dependence of angular resolution of the GAMMA-400 gamma-ray telescope in the energy range from 10 MeV to 3 TeV. PSF of the Fermi-LAT gamma-ray telescope for front configuration is also shown. (b) The GAMMA-400 energy resolution dependence from initial gamma energy (black line). The energy resolution of FERMI-LAT instrument for back configuration [6] (green line).}
\end{minipage}
\end{figure}

\section{GAMMA-400 ENERGY RESOLUTION}

The GAMMA-400 energy resolution dependence on the initial gamma energy is shown in FIG. 4b by black line. To calculate this value, only events with successful gamma-ray incident angle reconstruction were used. For each event the energy release in active matter of detectors was taken into account. Namely, the information from the position-sensitive silicon strips detectors of tracker C and the information from the deep electromagnetic calorimeter CC (FIG. 1) were used. Dotted line corresponds to the estimation of instrument limit due to the influence of electronic noises. The energy resolution of Fermi-LAT instrument with back configuration [6] is also presented in FIG. 4b by green line.
Involving some additional analysis of the conversion point position inside a converter-tracker system and taking into account the information from S1 and S2 layers, it is possible to significantly improve the energy resolution of gamma with energy less, than 100 MeV. For 50-MeV gamma-rays we obtained the value about 10$\%$.

\section{CONCLUSION}

To improve the incident angle reconstruction accuracy for low-energy gamma rays the following construction modification is used: in the converter-tracker C (FIG. 1) instead of 8 layers of tungsten with thickness 0.1X$_0$, 20 thin layers of converter foils with thickness 0.025X$_0$ are installed. Moreover, the special analysis of topology of pair-conversion events in thin layers of converter was performed. Choosing the pair-conversion events with more precise vertical localization allows us to improve the angular resolution. For 50-MeV gamma rays, the GAMMA-400 gamma-ray telescope angular resolution is 4.6$^{\circ}$ that is several degrees better than in the Fermi-LAT mission. The energy resolution for the considered physical scheme is $\sim$10$\%$ for 50-MeV gamma-rays.

\begin{acknowledgments}
This work was supported by National Research Nuclear University MEPhI in the framework of the Russian Academic Excellence Project (contract No. 02.a03.21.0005, 27.08.2013).
\end{acknowledgments}


\end{document}